\documentclass[showpacs,nopreprintnumbers,amsmath,pra,superscriptaddress]{revtex4}

\usepackage{graphicx}% Include figure files

\begin{document}

\title{Theory of dark resonances for alkali vapors in a buffer-gas cell}

\author{A. V. Ta\u\i chenachev}
\affiliation{Novosibirsk State University, Pirogova 2,
Novosibirsk 630090, Russia}
\affiliation{Institute of Laser Physics SD RAS,
Novosibirsk 630090, Russia}
\affiliation{Time and Frequency Division, NIST Boulder, 325 S. Broadway,
Boulder, CO 80305, USA}
\author{V. I. Yudin}
\affiliation{Novosibirsk State University, Pirogova 2,
Novosibirsk 630090, Russia}
\affiliation{Institute of Laser Physics SD RAS,
Novosibirsk 630090, Russia}
\affiliation{Time and Frequency Division, NIST Boulder, 325 S. Broadway,
Boulder, CO 80305, USA}
\author{R. Wynands}
\affiliation{Institut f\"ur Angewandte Physik, Universit\"at
Bonn, Wegelerstra\ss e 8, D-53115 Bonn, Germany}
\affiliation{Present address: D\'epartement de Physique, Universit\'e de
Fribourg, Chemin du Mus\'ee 3, 1700 Fribourg, Switzerland}
\author{M. St\"ahler}
\affiliation{Institut f\"ur Angewandte Physik, Universit\"at
Bonn, Wegelerstra\ss e 8, D-53115 Bonn, Germany}
\author{J. Kitching}
\affiliation{Time and Frequency Division, NIST Boulder, 325 S. Broadway,
Boulder, CO 80305, USA}
\author{L. Hollberg}
\affiliation{Time and Frequency Division, NIST Boulder, 325 S. Broadway,
Boulder, CO 80305, USA}

\date{\today}

\begin{abstract}
We develop an analytical theory of dark resonances that
accounts for the full atomic-level structure, as well as all field-induced
effects such as coherence preparation, optical pumping, ac Stark
shifts, and power broadening.  The analysis uses a
model based on relaxation constants that assumes the total
collisional depolarization of the excited state. A good qualitative
agreement with experiments for Cs in Ne is obtained.
\end{abstract}


\maketitle

\section{Introduction}

Nonlinear interference effects connected with the atomic ground state
coherence are now well known and widely used \cite{arimondo96}.  One of
the most promising classes of these effects, especially for precise
measurements, is that of super-narrow dark resonances
\cite{wynands99,kitching00,marcus01} that appear in the medium's
response to bichromatic laser excitation, when the laser frequency
difference is close to the atomic ground-state splitting.  The use of
vapor cells containing a buffer gas in addition to an alkali vapor has
allowed the measurement of resonance linewidths less than 50 Hz
\cite{brandt96,erhard01}. While such resonances have been extensively
investigated experimentally (especially in the case of Cs)
\cite{wynands99}, a detailed theoretical understanding is not yet well
developed for realistic multilevel systems, motivating the present work.
Our theory was developed in close connection with ongoing efforts to
construct compact atomic clocks
\cite{kitching00,knappe01,kitching01,kitching02} and magnetometers
\cite{wynands99,marcus01}. For any practical application of dark
resonances, the stability and accuracy are optimized with respect to
parameters such as the output signal amplitude, width, and shift. In the
problem considered here, many parameters, such as laser detunings, field
component polarizations and amplitudes, and buffer gas pressure, affect
the dark resonance itself. In addition, various excitation schemes (for
example, $D_2$ versus $D_1$ line excitation \cite{kitching02b}) and
different atomic isotopes can be used. A natural question arises: what
design will optimize the performance of the clock (or magnetometer)?
Previous theories did not completely answer this question. One main
obstacle was connected with the complicated energy-level structure of
the real atomic systems used in experiments.

Generally speaking, there are several types of problems in the
theoretical description of dark resonances. One problem relates to a
proper treatment of the relaxation processes in the system, including
velocity-changing collisions \cite{arimondo96vcc} and the spatial
diffusion of coherently prepared atoms \cite{vanier89,zibrov01}. Light
propagation through coherently prepared nonlinear media, especially
through optically thick media \cite{lukin97}, can be thought of as
another type of difficulty.  This paper addresses another important
problem: that of field-induced processes in multilevel systems such as
coherence preparation, optical pumping, ac Stark shifts, and power
broadening. All existing theories can be classified into three kinds:
few-state models (basically, three-state lambda systems)
\cite{ling96,grishanin98,erhard01}, perturbation theories
\cite{wynands98}, and numerical simulations \cite{ling96,erhard01}. All
three classes of theories have disadvantages. The first theory neglects
many details of the actual configuration of atomic levels. Perturbation
theory neglects some effects induced by the presence of the optical
field (namely, optical pumping, ac Stark shifts, and power broadening).
Numerical simulation theories demonstrate a lack of genuine
understanding and predictive power.

This paper presents a new analytical theory, that accounts for the level
structure (both Zeeman and hyperfine) of a real atom, as well as all
field-induced effects. The relaxation processes are treated in the
simplest way: by neglecting velocity-changing collisions and all effects
connected with the spatial inhomogeneity, we reduce the model to one
described simply by relaxation constants. The crucial assumption is
total collisional depolarization of the excited state. In addition, we
add the (optional) approximations of homogeneous broadening and low
saturation.  With these approximations, a general analytical result is
obtained for the atomic response, which result is valid for arbitrary
excitation schemes ($D_2$ as well as $D_1$ lines), light field
polarizations, and magnetic fields. In the specific case of circularly
polarized light in the presence of a magnetic field, where only two
states participate in the coherence preparation, analytical lineshapes
(generalized Lorentzian) coincide exactly with the phenomenological
model heuristically introduced previously to fit experimental data
\cite{knappe02}. In the case of zero magnetic field, and when
contributions of different Zeeman sub-states are well overlapped, the
resonance lineshape is also approximately described by the generalized
Lorentzian. A comparison of analytically calculated coefficients of the
Lorentz-Lorenz model (with no free parameters) with coefficients
extracted from experimental data demonstrates a good qualitative
agreement.

\section{Statement of the problem}

\begin{figure}
\includegraphics[width=3in]{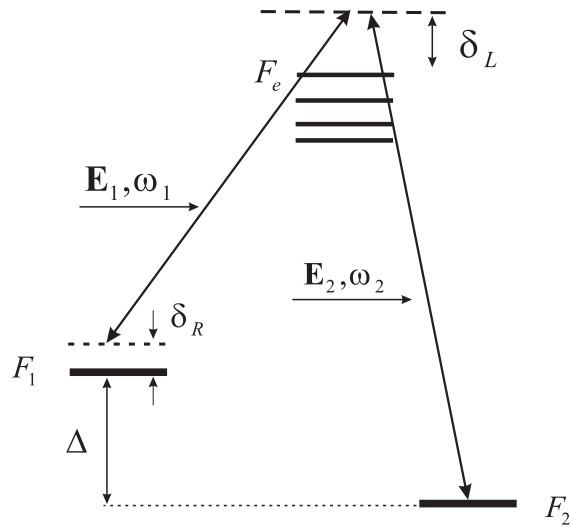}
\caption{Excitation scheme}
\label{fig1}
\end{figure}

In this section, the general framework of the problem is described, the
basic assumptions we make are stated and the specific procedure for
calculating the quantities of interest is outlined. We consider the
resonant interaction of alkali atoms in the $S_{1/2}$ ground state with
a two-frequency laser field
\begin{equation} \label{2frfield}
{\bf E}(z,t) = {\bf E}_1 \exp[-i(\omega_1 t-k_1 z)] +
{\bf E}_2 \exp[-i(\omega_2 t-k_2 z)]  + c.c.
\;,
\end{equation}
where both components propagate in the positive direction ($k_{1,2} >0$).
The field can excite atoms either to the $P_{1/2}$ state ($D_1$ line) or
to the $P_{3/2}$ state ($D_2$ line). Two hyperfine (HF)
components are present in the ground state with the total angular momenta $F_1 =
I+1/2$ and  $F_2 = I-1/2$ (where $I$ is the nuclear spin). The HF splitting in
the ground state $\Delta = ({\cal E}_1 - {\cal E}_2)/\hbar$ is in the
range $1$ to $10\,$GHz. The excited state has two ($D_1$ line) or four ($D_2$
line) HF levels with the angular momenta $F_e = I-J_e,\ldots,I+J_e$ and
the energies ${\cal E}_e = \hbar \omega_e$. The HF splitting of the
excited state is typically one order of magnitude smaller than~$\Delta$.
 To be more specific, we assume that the frequency $\omega_1$ is close
to resonance with the $F_1 \to F_e$ transitions, while  the other
frequency $\omega_2$ is close to the frequencies of the $F_2 \to F_e$
transitions.  Thus, we have a $\Lambda$-type excitation scheme
(Fig.~\ref{fig1}). In the absence of an external B-field, the HF levels
are degenerate with respect to the total angular momentum projections.
For the Zeeman sub-states the following shorthand notations will be used:
$|e\rangle = |F_e,\,m_e\rangle$ with $m_e = -F_e,\ldots,F_e$, and
$|i,m\rangle = |F_i,\,m\rangle$ with $m = -F_i,\ldots,F_i$ ($i=1,2$).

For simplicity, we consider first an atom at rest, positioned at the origin $z=0$.
Each frequency component of the field can in principal induce
transitions from both ground-state HF levels. Then the interaction
Hamiltonian in the dipole approximation contains contributions of two
kinds:
\begin{equation} \label{HDE}
\widehat{H}_{D-E} = -
\sum_{e,i,m} |e\rangle \langle e|
(\widehat{\bf d}\cdot {\bf E}_i)|i,m\rangle \langle i,m|-
\sum_{e,i\neq j,m} |e\rangle \langle e|
(\widehat{\bf d}\cdot {\bf E}_i)|j,m\rangle \langle j,m|
e^{-i(\omega_i-\omega_j)t}
+h.c. \;,
\end{equation}
where we use a rotating frame (the unitary transformation of the
ground-state basis $|i,m\rangle \to \exp(i\omega_i
t)|i,m\rangle$), and $\widehat{\bf d}$ is the dipole moment
operator. The first term in \eqref{HDE} is independent of time in
the rotating basis, and we refer to it as the resonant
contribution. The second term, oscillating at the difference
frequency, results in off-resonant contributions to the  optical
shifts and optical pumping rates, as well as in temporal
oscillations of the atomic density matrix. The role of the
off-resonant term in the case of a three-level $\Lambda$ system
has been studied in great detail~\cite{grishanin98}. The
amplitudes of the oscillating parts of the density matrix can be
approximated as $|dE|^2/(\hbar \Delta)^2$. For the moderate field
intensities considered here ($< 10\,$mW/cm$^2$) this ratio is very
small, $|dE|^2/(\hbar \Delta)^2 \sim 10^{-6}-10^{-8}$, and the
oscillating terms can be safely neglected. However, the
off-resonant contributions to the optical energy shifts and widths
can be significant, especially in the case of large one-photon
detunings. \label{oscill}

The Hamiltonian for a free atom in the rotating frame can be written as
\begin{equation} \label{H0}
\widehat{H}_0 = - \sum_{e}\hbar(\delta_L-\omega_e)|e\rangle \langle e|
- \hbar \frac{\delta_{R}}2
\sum_{m}(|1,m\rangle \langle 1,m| -
|2,m\rangle \langle 2,m|)  \;.
\end{equation}
Here $\delta_L = (\delta_1+\delta_2)/2$ is the average one-photon
detuning, $\delta_L$ and $\omega_e$ are measured from a common zero
level (for example, from the HF level with maximal momentum $F_e =
I+J_e$), and $\delta_R = \delta_2-\delta_1 = \omega_2-\omega_1 -\Delta$
is the Raman (two-photon) detuning.

Since this paper is concerned with the field-induced effects
in multi-level atomic systems, the relaxation processes are modeled
by several constants. The homogeneous broadening of the optical line,
due mainly to collisions with a buffer gas, is described by the
constant $\gamma$. We assume that the excited state is completely
depolarized due to collisions during the radiative lifetime $\tau_e$,
i.e., the depolarization rates $\gamma_{\kappa}$ obey the condition
\begin{equation} \label{depol_cond}
\gamma_{\kappa}\,\tau_e \gg 1 \;.
\end{equation}
The relaxation of the ground-state density matrix to the isotropic
equilibrium, both due to the diffusion through the laser beam
and due to collisions, is modeled by a single constant $\Gamma$.

Under the assumption of moderate field intensities and high buffer-gas
pressure, we develop the theory in the low-saturation limit:
\begin{equation} \label{lowsat}
\frac{|dE|^2}{\hbar^2} \ll \frac{\gamma}{\tau_e} \;.
\end{equation}
The two-photon dark resonance appears when the Raman detuning $\delta_R$
is scanned around zero. The width of the dark resonance, which is related to
the ground-state relaxation, is usually six orders of magnitude smaller
than the optical linewidth $\gamma$. The approximation $\delta_R
\ll \gamma$ is therefore suitable.

It should be stressed that all approximations are well justified for
typical experimental conditions. For example, in the case of Cs in a
background Ne atmosphere at a pressure of $p = 10\,$kPa, the homogeneous
broadening $\gamma \approx 2\pi\,860\,$MHz \cite{allard82} of the
optical line exceeds the Doppler width $k\overline{v} \approx
2\pi\,300\,$MHz, so velocity-changing collisions are inconsequential. The
collisional depolarization rate $\gamma_{\kappa} \approx 2\pi\,70\,$MHz
\cite{happer72} is large compared to the inverse radiative lifetime
$1/\tau_e = 2\pi\,5.3\,$MHz. The Rabi frequency $|dE|/\hbar \approx
1/\tau_e$ for the field intensity $8.8\,$mW/cm$^2$, which results in a
saturation parameter $(|dE|/\hbar)^2 \,\tau_e/\gamma \approx 10^{-2}$.
The two-photon detuning is scanned in the range $|\delta_R| <
2\pi\,1\,$MHz, and the ground-state relaxation rate can be estimated to be
$\Gamma \approx 2\pi\, 53\,$Hz \cite{beverini71,vanier89}.

Eliminating optical coherences with these approximations (for
details see the Appendix), we arrive at the following set of
equations for the ground-state density submatrix
($\widehat{\sigma}_{gg} =
\widehat{\Pi}_g\widehat{\sigma}\widehat{\Pi}_g$):
\begin{eqnarray} \label{maineq}
&& \frac{d}{dt}\widehat{\sigma}_{gg} =
-i\left[ \widehat{H}_{\rm eff}\widehat{\sigma}_{gg} -
\widehat{\sigma}_{gg}{\widehat{H}_{\rm eff}}^{\dagger} \right] +
\left(\frac{\pi_e}{\tau_e}+\Gamma\right)\,\frac{\widehat{\Pi}_g}{n_g},
\\
\label{n_cond}
&& {\rm Tr}\{\widehat{\sigma}_{gg}\} = 1 \;,
\end{eqnarray}
where $\widehat{\Pi}_g = \sum_{m}(|1,m\rangle \langle 1,m| + |2,m\rangle
\langle 2,m|)$ is the ground-state projector, $n_g=2(2I+1)$ is the total
number of sub-states in the ground state, and $\pi_e$ is the total
population of the excited state. The first term ($\propto \pi_e$) of the
source in \eqref{maineq} corresponds to the isotropic repopulation of
the ground-state sublevels due to the spontaneous decay of the excited
states. The other term ($\propto \Gamma$) describes the entrance of
unpolarized atoms due to diffusion and collisions. Due to the
conservation of the total number of particles (\ref{n_cond}), separate
dynamic equations for the excited-state density matrix elements are not
needed. Both the dynamics and steady state are completely governed by
the non-Hermitian ground-state Hamiltonian:
\begin{equation} \label{Heff}
\widehat{H}_{\rm eff} = -\frac{\delta_{R}}2
\sum_{m}(|1,m\rangle \langle 1,m| -
|2,m\rangle \langle 2,m|)
 + \widehat{R} -i\frac{\Gamma}{2}\,\widehat{\Pi}_g\;.
\end{equation}
Here the excitation matrix,
\begin{eqnarray} \label{Rmatr}
\widehat{R} &=& \sum_{i,j,e,m,m'}|i,m\rangle
\frac{\langle i,m|{(\widehat{\bf d}\cdot {\bf
E}_i)}^{\dagger}|e\rangle \langle e| (\widehat{\bf
d}\cdot {\bf E}_j)|j,m'\rangle}
{\hbar^2\,[(\delta_L-\omega_e)+i\gamma/2]} \langle j,m'| +
\nonumber \\
&+&
\sum_{i\neq j,e,m,m'}|i,m\rangle
\frac{\langle i,m|{(\widehat{\bf d}\cdot {\bf
E}_j)}^{\dagger}|e\rangle \langle e| (\widehat{\bf
d}\cdot {\bf E}_j)|i,m'\rangle}
{\hbar^2\,[(\delta_L+\omega_j-\omega_i-\omega_e)+i\gamma/2]}
\langle i,m'|,
\end{eqnarray}
contains the resonant (first summation) as well as off-resonant (second
summation) contributions to the optical shifts and optical
pumping rates (Hermitian and anti-Hermitian parts, respectively). The
non-diagonal ($i\neq j$) elements of the resonant term induce the Raman
coherence between the HF levels of the ground state responsible for the
dark resonance.

The generic matrix element in (\ref{Rmatr}) is calculated
from the Wigner-Eckart theorem:
\begin{eqnarray} \label{generic}
&& \langle i,m_i|{(\widehat{\bf d}\cdot {\bf
E}_k)}^{\dagger}|e\rangle \langle e| (\widehat{\bf
d}\cdot {\bf E}_l)|j,m_j\rangle =
|\langle J_e||d||J_g \rangle|^2\, r(F_e,F_i)\, r(F_e,F_j) \times
\nonumber \\
&& \times\,
\sum_{K,\,q} (-1)^{F_e+F_j+K}
\left\{
\begin{array}{rcl}
1 & 1 & K\\
F_i & F_j & F_e
\end{array}
\right\}
\sqrt{2K+1}\,(-1)^{F_i-m_i}
\left(
\begin{array}{rcl}
F_i & K & F_j\\
-m_i & q & m_j
\end{array}
\right)
\{{{\bf E}_k}^* \otimes {\bf E}_l\}_{K\,q} \;,
\end{eqnarray}
where
% $J_g$ and $J_e$ denote the electronic angular momentum in the
% ground and excited states;
$\langle J_e||d||J_g \rangle$ is the
reduced matrix element of the dipole moment and
\[
r(F_e,F_i) = \sqrt{(2J_e+1)(2F_e+1)(2F_i+1)} \,
\left\{
\begin{array}{rcl}
J_g & J_e & 1\\
F_e & F_i & I
\end{array}
\right\}
\]
is the partial coupling amplitude of the $F_i \to F_e$ transition. In
the general case we have scalar ($K=0$), vector ($K=1$) and quadrupole
($K=2$) contributions. All possible selection rules are contained in the
coefficients of vector coupling, i.e., the $6j$ and $3jm$ symbols.

For an atom moving along the direction of propagation of the optical
field, the field frequencies are shifted due to the Doppler effect:
$\omega_i \to \omega_i-k_i v$. As a result, a Doppler shift of the
one-photon detuning $\delta_L \to \delta_L -kv$ occurs, where
$k=(k_1+k_2)/2$, as does a residual Doppler shift of the Raman detuning
$\delta_R \to \delta_R -(k_2-k_1)v$. At high buffer-gas pressure the
residual Doppler shift is suppressed due to the Lamb-Dicke effect
\cite{dicke53,vanier89}. However, in the general case the Doppler shift
of the one-photon detuning can be significant, and certain quantities
must be averaged over the Maxwell velocity distribution. Nevertheless,
for buffer-gas pressures typically used in experiments, the
approximation of homogeneous broadening is reasonable as a first
approach to the problem, because the homogeneous width $\gamma$ equals
or even exceeds the Doppler width $k \overline{v}$.

Here we consider the steady-state regime, setting
$(d/dt)\,\widehat{\sigma}_{gg} = 0$ in \eqref{maineq}. As a
spectroscopic signal, we consider the total excited-state population
$\pi_e$, which is proportional to the total light absorption in
optically thin media or to the total fluorescence. The following
procedure is used to find $\pi_e$. From \eqref{maineq}, the ground-state
density matrix $\widehat{\sigma}_{gg}$ is expressed in terms of $\pi_e$,
and then $\pi_e$ is calculated from the normalization condition
\eqref{n_cond}. The solution of this algebraic problem can be obtained
in a compact analytical form in two important special cases. The first
arises when both field components have the same simple (circular or
linear) polarization and there is no magnetic field. Here, for a
suitable choice of the quantization axis, the excitation matrix
$\widehat{R}$ contains only diagonal elements with respect to the
magnetic quantum number, i.e., $m=m'$ in \eqref{Rmatr}. The second case
appears when a magnetic field is applied and just a few substates
contribute to the Raman coherence for arbitrary light
polarizations and arbitrary magnetic field directions. Both cases are
considered below.

\section{Simple light polarization, no magnetic field}

\begin{figure}
\includegraphics[width=2in]{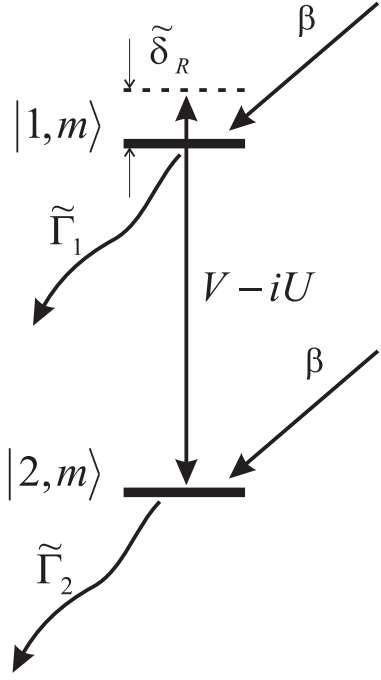}
\caption{Effective two-level system, corresponding to one $m$-block}
\label{fig2}
\end{figure}

We turn now to the case of circular field polarization when the
quantization axis is directed orthogonal to the polarization vector (or
alternatively linear polarization when the quantization axis is aligned
along the polarization vector). We evaluate the total excited-state
population, $\pi_e$, in order to determine how the dark resonance signal
(proportional to $\pi_e$) depends on parameters such as the optical
detuning from resonance. Under these assumptions, the complete set of
equations (\ref{maineq}) can be split into independent blocks for each
magnetic quantum number $m$ ($m$-blocks). These blocks for $m=\pm F_1$
contain only one equation for the sub-state population $\pi^{(\pm
F_1)}$. The other blocks with $m \neq \pm F_1$ contain four equations
(two for the populations and two for the Raman coherences),
corresponding to an effective two-level system with the upper $|1,m
\rangle$ and lower $|2,m \rangle$ states (Fig.~\ref{fig2}). The
parameters of the two-level system are expressed in terms of matrix
elements of $\widehat{R}$ as follows: the population relaxation rates
$\widetilde{\Gamma}_i = \Gamma + R^{(m)}_i$  include the optical pumping
rates $R^{(m)}_i = 2\,{\rm Im}\{\langle i,m|\widehat{R}|i,m\rangle\}$;
the dephasing rate is $\widetilde{\Gamma}_{12} =
(\widetilde{\Gamma}_1+\widetilde{\Gamma}_2)/2$; the effective detuning
$\widetilde{\delta}_R = \delta_R-(S^{(m)}_1-S^{(m)}_2)$ includes optical
shifts $S^{(m)}_i = {\rm Re}\{\langle i,m|\widehat{R}|i,m\rangle\}$; and
the coherence between levels is excited by the complex coupling $V-iU =
\langle 1,m|\widehat{R}|2,m\rangle$. Note that the phase of the matrix
element $\langle 1,m|{(\widehat{\bf d}\cdot {\bf
E}_1)}^{\dagger}|e\rangle \langle e| (\widehat{\bf d}\cdot {\bf
E}_2)|2,m\rangle $ can be chosen equal to zero without loss of
generality, so that $\langle 2,m|\widehat{R}|1,m\rangle=\langle
1,m|\widehat{R}|2,m\rangle$.

Both the upper and lower states are repopulated with the same rate
$\beta = (\pi_e/\tau_e+\Gamma)/n_g$. First the total
$m$-block population $\pi^{(m)} = \pi^{(m)}_1+\pi^{(m)}_2$ per
unit repopulation rate is found. For the outermost blocks, $m=\pm F_1$,
$\pi^{(\pm F_1)} = 1/(\Gamma+R^{(\pm F_1)}_1)$.
The result for $m\neq \pm F_1$ is a quotient of
polynomials of second order in the effective detuning,
\begin{equation} \label{pim}
\pi^{(m)} = \frac{A\, \widetilde{\delta}_R^2+ B}{
C\,\widetilde{\delta}_R^2+D\, \widetilde{\delta}_R+E} \;,
\end{equation}
where
\begin{eqnarray} \label{ABCDE}
A &=& \widetilde{\Gamma}_1+\widetilde{\Gamma}_2
\;;\;\;\;\;\;
B =
\widetilde{\Gamma}_{12}
[\widetilde{\Gamma}_{12}(\widetilde{\Gamma}_1
+\widetilde{\Gamma}_2)+8\,V^2]
\;;
\nonumber \\
C &=& \widetilde{\Gamma}_1 \widetilde{\Gamma}_2
\;;\;\;\;\;\;\;\;\;\;
D = 4\,UV(\widetilde{\Gamma}_1-\widetilde{\Gamma}_2) \;;
\\
E &=& \widetilde{\Gamma}_{12}^2 \widetilde{\Gamma}_1
\widetilde{\Gamma}_2 + 2\,
\widetilde{\Gamma}_{12}(\widetilde{\Gamma}_1+
\widetilde{\Gamma}_2)(V^2-U^2) - 16\,U^2 V^2
\nonumber \;.
\end{eqnarray}
The repopulation rate, corresponding to unit total population in
all $m$-blocks is
\begin{equation} \label{beta}
\beta = \left[\sum_{m=-F_1}^{F_1}\pi^{(m)}\right]^{-1} \;,
\end{equation}
and the total excited-state population is finally expressed as
\begin{equation} \label{pie}
\pi_e = \tau_e(n_g\,\beta-\Gamma) \;.
\end{equation}

In the general case, when polarizations of the field components are
different, or the same but elliptical, there is no basis where the
matrices $\langle 1,m|\widehat{R}|1,m'\rangle$, $\langle
2,m|\widehat{R}|2,m'\rangle$, and $\langle 1,m|\widehat{R}|2,m'\rangle$
are simultaneously diagonal. In this situation, the full equation set
for the ground-state density matrix elements must be solved, including
all possible Zeeman and Raman coherences. Nevertheless, one important
exception should be noted. If the optical linewidth is much greater than
the excited-state HF splitting $\gamma \gg
(\omega_{e,\,\mathrm{max}}-\omega_{e,\,\mathrm{min}})$, the quadrupole
contributions to $\widehat{R}$ are negligible \cite{wynands98}. The
vector terms are diagonal (with respect to the magnetic quantum number)
in the coordinate frame with $z$ as the quantization axis, since $[{{\bf
E}_i}^*\times{\bf E}_j] \propto {\bf e}_z$. Thus, we return to the case
discussed above.

\section{Dark resonances in a magnetic field}

In a weak magnetic field, the ground-state magnetic sublevels are
split due to the linear Zeeman effect, which can be described by
the following additional term in the effective Hamiltonian
(\ref{Heff}):
\begin{equation} \label{HB}
\widehat{H}_{B} = \sum_{i,m} m\,\Omega_i\,
|i,m\rangle  \langle i,m| \;.
\end{equation}
Here the quantization axis is directed along the magnetic field, and
$\Omega_i = \mu_B g_i B/\hbar$ are the Zeeman splitting frequencies,
with $\mu_B$ the Bohr magneton and $B$ the magnetic flux density. The
$g$-factors of levels, $g_i$, are expressed through the electronic
$g_{J}$ and nuclear $g_{I}$ Lande factors:
$$
g_{1,2} = \pm\, \frac{g_J-g_I}{2\,I+1}+g_I \;.
$$
The magnetic field causes a precession of atomic coherences with
frequencies $m\,\Omega_i -m'\,\Omega_j$. When the Zeeman frequencies are
much larger than off-diagonal elements of the excitation matrix
$\Omega_i \gg |\langle i,m|\widehat{R}|i,m'\rangle|$, the light-induced
Zeeman coherences within the $i$-th HF level are negligible. Thus, we
again have a set of independent two-level systems, consisting of the
sub-states $|1,m_1\rangle$ and $|2,m_2\rangle$ (where $|m_1-m_2| \leq 2$
due to the selection rules). The formulas (\ref{pim}) and (\ref{ABCDE})
for the total block population are still valid for every
$(m_1,\,m_2)$-block with the following substitutions:
\begin{eqnarray} \label{m1m2subs}
\widetilde{\Gamma}_i &=& \Gamma + R^{(m_i)}_i
\;;\;\;\;
\widetilde{\delta}_R = \delta_R -(S_1^{(m_1)}-S_2^{(m_2)}) -(m_1
\Omega_1 - m_2 \Omega_2)\;;
\nonumber\\
V &-& i U = \langle 1,m_1|\widehat{R}|2,m_2\rangle
= \langle 2,m_2|\widehat{R}|1,m_1\rangle
\;.
\end{eqnarray}

If the Zeeman frequencies significantly exceed the widths
$\widetilde{\Gamma}_i$, the Zeeman-split dark resonances are well
resolved. In other words, the Raman coherence between the substates
$|1,m_1\rangle$ and $|2,m_2\rangle$ is effectively induced  when the
precession frequency is approximately equal to the Raman detuning:
$\delta_R \approx m_1\,\Omega_1 - m_2\,\Omega_2$. This condition can be
simultaneously satisfied for only a few $(m_1,\,m_2)$-blocks. More
precisely, the nuclear Lande factor is typically three orders of
magnitude smaller than the electronic Lande factor (for cesium $g_J/g_I
\approx 2500$); then, with good accuracy, $\Omega_1 = - \Omega_2
=\Omega$ and the Zeeman shift of the dark resonance position is
proportional to the sum of magnetic quantum numbers $n\Omega=(m_1\,+
m_2)\,\Omega$. It can be seen that, in the general case, three blocks
$(m,\,m)$, $(m-1,\,m+1)$, and $(m+1,\,m-1)$ contribute to the coherence
preparation for the resonances with even shifts $2\,m\,\Omega$, and two
other blocks $(m-1,\,m)$ and $(m,\,m-1)$ contribute for the resonances
with odd shifts $(2\,m-1)\,\Omega$. When $\delta_R$ is tuned around the
resonance with given shift $n\,\Omega$, the repopulation rate $\beta$
can be written as
\[
\beta = \left[Z+\sum_{m_1+m_2 = n}
\pi^{(m_1,m_2)}(\widetilde{\delta}_R)
\right]^{-1} \;,
\]
where the first summand $Z$ does not depend on the Raman detuning:
\[
Z=\sum_{m_1+m_2 \neq n}  \left( \frac{1}{\Gamma+R^{(m_1)}_1}+
\frac{1}{\Gamma+R^{(m_2)}_2}\right) \;,
\]
and $\pi^{(m_1,m_2)}$ is the total population of the $(m_1,\,m_2)$
block. Owing to the nuclear contribution, a further increase of the
magnetic field causes the dark resonances to be eventually split into
individual peaks, corresponding to each $(m_1,\,m_2)$-block
\cite{knappe99}.

\section{The resonance lineshape}

We now consider the dark resonance lineshape in more detail.  First,
we analyze the particular case in which just two sub-states
$|1,\,0\rangle$ and $|2,\,0\rangle$ participate in the Raman
coherence, i.e., we consider the magnetically insensitive resonance
($m=0$) in a magnetic field. This $(0,\,0)$ resonance is of primary
interest for possible clock applications
\cite{wynands99,kitching00,knappe01}, because it is only sensitive to
a magnetic field in second order. Here the absorption signal,
$n''_{DR}$, has the form:
\begin{equation} \label{00abs}
n''_{DR} = \frac{\pi_e}{\tau_e\,n_g} =
\frac {1}{Z+\pi^{(0)}(\widetilde{\delta}_R)} -
\frac{\Gamma}{n_g}
\;;\;\;\;\;
Z = \sum_{m \neq 0}  \left( \frac{1}{\Gamma+R^{(m)}_1}+
\frac{1}{\Gamma+R^{(m)}_2}\right) \;,
\end{equation}
where $\pi^{(0)}$ is the total population of the $(m=0)$-block
per unit repopulation rate (see \eqref{pim} and \eqref{ABCDE}).
Since $\pi^{(0)}$ is a quotient of polynomials of second order in
$\delta_R$, the absorption can be written as the sum of an absorptive
and a dispersive Lorentzian, and a constant background:
\begin{equation} \label{LLc}
n''_{DR} = -C_1\,\frac{(\widetilde{\gamma}/2)^2}
{(\widetilde{\gamma}/2)^2+(\delta_R-\delta_0)^2} +
C_2\,\frac{(\delta_R-\delta_0)\,\widetilde{\gamma}/2}
{(\widetilde{\gamma}/2)^2+(\delta_R-\delta_0)^2} + \mathrm{const}
\;.
\end{equation}
The parameters in \eqref{LLc} are expressed in terms of the coefficients
introduced by \eqref{ABCDE} in the following way. The dark resonance
position is governed by the optical shifts and an additional term caused
by the two-photon coupling between levels:
\begin{equation} \label{d0}
\delta_0 = (S_1^{(0)}-S_2^{(0)})+x \;;\;\;\;\;
x=-\frac{D Z}{2\,(A+C Z)} \;.
\end{equation}
The width of dark resonance reads
\begin{equation} \label{width}
(\widetilde{\gamma}/2)^2 = \frac{B+E Z}{A+C Z} - x^2  \;.
\end{equation}
The amplitudes of the symmetrical and antisymmetrical Lorentzians are
found from the relations
\begin{eqnarray} \label{c1}
C_1 (\widetilde{\gamma}/2)^2 &=& \frac{B C - A E -x A D}{(A+C Z)^2}
\;;\\
\label{c2}
C_2\, \widetilde{\gamma}/2 &=& \frac{A D}{(A+C Z)^2}
\;.
\end{eqnarray}
The background constant, $C/(C Z+A)-\Gamma/n_g$, corresponds to the
absorption far off the two-photon resonance.

The result \eqref{LLc} for the resonance
lineshape is quite general. In fact, it does not depend on
our simplified assumptions on the relaxation processes but is valid also
in the low-saturation limit for arbitrary relaxation matrix,
whenever only two states participate in the coherence preparation and
$\delta_R \ll \gamma$.

Turning to the case of zero magnetic field and simple field
polarization, we proceed with the goal of determining the resonance
position, width and amplitudes of the symmetrical and asymmetrical
components as above. Since all Zeeman levels within a given hyperfine
level are now degenerate, we rewrite the repopulation rate $\beta$
\eqref{beta} as:
\begin{equation} \label{betar}
\beta = \left[
Z + (2F_2+1) \langle \widetilde{\pi}^{(m)}(\delta_R)\rangle_m
\right]^{-1} \;,
\end{equation}
where
$$
Z = \sum_{m =-F_1}^{m=F_1}  \left( \frac{1}{\Gamma+R^{(m)}_1}+
\frac{1}{\Gamma+R^{(m)}_2}\right)
$$
does not depend on $\delta_R$ and corresponds to the absorption far
off the two-photon resonance; the sum of the variable parts of the
$m$-block populations $\widetilde{\pi}^{(m)}(\delta_R)$ is expressed through the
average over $m$-blocks, where the average of a variable $X$ is defined as:
$$
\langle X^{(m)}\rangle_m = \frac 1{2F_2+1}
\sum_{m =-F_2}^{m=F_2} X^{(m)} \;.
$$
Since $\widetilde{\pi}^{(m)}(\delta_R)$ is a quotient of polynomials of second
order:
\begin{eqnarray} \label{mum}
&& \widetilde{\pi}^{(m)}(\delta_R) = \frac{a_2^{(m)}\delta_R
+b_2^{(m)}}{\delta_R^2+a_1^{(m)}\delta_R+ b_1^{(m)}} \;;
\nonumber\\
&& a_1^{(m)} = \frac{D}{C} -(S_1^{(m)}-S_2^{(m)}) \;;\;\;\;
b_1^{(m)} = \frac{E}{C} - (S_1^{(m)}-S_2^{(m)})\frac{D}{C}+
(S_1^{(m)}-S_2^{(m)})^2 \; ;
\nonumber\\
&& a_2^{(m)} = \frac{A\,D}{C^2}  \;;\;\;\;
b_2^{(m)} = \frac{B\,C-A\,E+A\,D\,(S_1^{(m)}-S_2^{(m)})}{C^2} \;,
\end{eqnarray}
the average $\langle \widetilde{\pi}^{(m)}(\delta_R)\rangle_m$ is
a quotient of polynomials of order $2\,(2F_2+1)$.
Generally this average describes a superposition of
resonances with different widths and positions due to the
$m$-dependent power broadening and ac Stark shifts, but if the laser
detuning is not too large, $|\delta_L| \leq \Delta$, all resonances are
well overlapped, and the average
$\langle\,\widetilde{\pi}^{(m)}(\delta_R)\rangle_m$ can be approximated by a
quotient of polynomials of second order. Here we use the following
simple procedure, where the average of a quotient is substituted by a
quotient of the averages:
\begin{equation}
\label{mumapp}
\langle \widetilde{\pi}^{(m)}(\delta_R)\rangle_m \approx \alpha
\frac{\langle a_2^{(m)}\rangle_m \delta_R
+\langle b_2^{(m)}\rangle_m}{\delta_R^2+\langle
a_1^{(m)}\rangle_m \delta_R+ \langle b_1^{(m)}\rangle_m} \;,
\end{equation}
and where the correction factor $\alpha$ is chosen such that the exact and
approximate expressions coincide at $\delta_R = 0$, i.e.,
$$
\alpha = \frac{\langle b_1^{(m)}\rangle_m}{\langle
b_2^{(m)}\rangle_m} \,\left\langle \frac{b_2^{(m)}}{b_1^{(m)}}
\right\rangle_m \;.
$$
Our approximation for $\beta$ yields an error less
than a few percent across a wide range of parameters.
With this approximation, we return to the resonance lineshape
(\ref{LLc}), where the parameters are expressed in terms of the
averages over $m$:
\begin{eqnarray} \label{LLpars}
\delta_0 &=& -\frac{\langle
a_1^{(m)}\rangle_m}{2}-\frac{(2F_2+1)\,\alpha}{Z}\,
\frac{\langle a_2^{(m)}\rangle_m}{2}
\nonumber\\
(\widetilde{\gamma}/2)^2 &=& \langle b_1^{(m)}\rangle_m +
\frac{(2F_2+1)\,\alpha}{Z}\,\langle b_2^{(m)}\rangle_m -\delta_0^2
\nonumber\\
C_1\,(\widetilde{\gamma}/2)^2 &=& (2F_2+1)\,\alpha\,
\frac{\langle b_2^{(m)}\rangle_m + \langle a_2^{(m)}\rangle_m
\delta_0}{Z^2}
\nonumber\\
C_2\,(\widetilde{\gamma}/2) &=& (2F_2+1)\,\alpha\,
\frac{\langle a_2^{(m)}\rangle_m }{Z^2}
\nonumber\\
\mathrm{const} &=& \frac{1}{Z} -\frac{\Gamma}{n_g} \;.
\end{eqnarray}

\section{Comparison with experiment}

The analytical lineshape \eqref{LLc} coincides exactly with the
phenomenological model heuristically introduced previously to fit
experimental data \cite{knappe02}. In those experiments a
vertical-cavity surface-emitting laser (VCSEL) was modulated at
the 9.2-GHz hyperfine splitting frequency of the cesium atom,
so that the laser output spectrum contained modulation sidebands
at this frequency. Using the carrier and one of the sidebands
the dark resonance could be prepared and spectroscopically
observed, as a function of the detuning~$\delta_L$ of the laser frequency
from optical resonance. Data was taken for three different power ratios
of carrier and sideband, with the cesium atoms contained
in a cell with 8.7\,kPa of neon as a buffer gas. Detection used
a modulation technique that allowed to extract simultaneously
the absorption and the dispersion line shape \cite{wyn99}.
For each detuning~$\delta_L$, both line shapes were simultaneously
fitted by the model function~\eqref{LLc}, with $C_1$, $C_2$, $\widetilde\gamma$,
and~$\delta_0$ as free parameters. Actually, as far as the line shapes
themselves are concerned, this is a two-parameter fit: $C_2/C_1$
and $\widetilde\gamma$ describe the shape, and the rest the
overall amplitude and position of the dark line.

\begin{figure}
\includegraphics[width=3in]{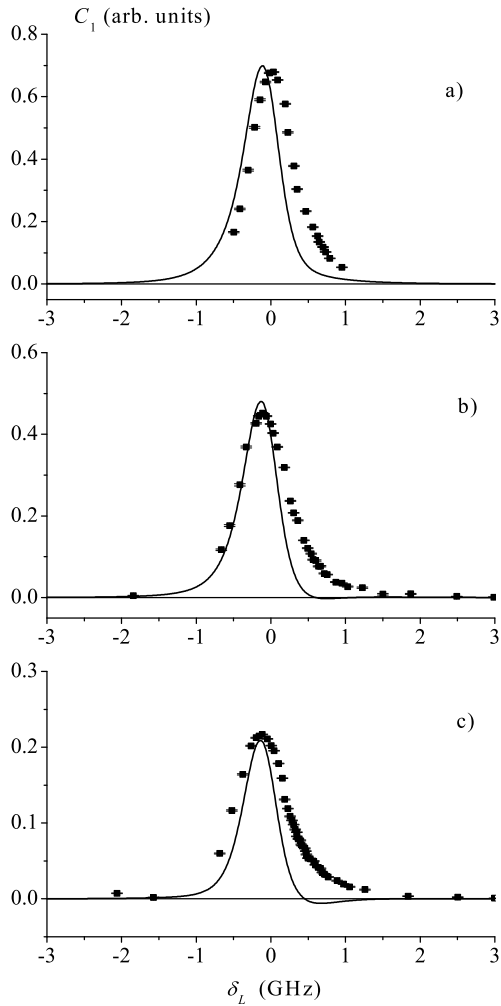}
\caption{Absorptive coefficient, $C_1$, versus optical detuning,
$\delta_L$. Plots a), b) and c) are for ${\cal R} = 2.4,\,7.2,\,22$,
respectively. The solid lines indicate the theoretical predictions while
the points indicate the experimental data taken from~\protect\cite{knappe02}.}
\label{fig3}
\end{figure}

\begin{figure}
\includegraphics[width=3in]{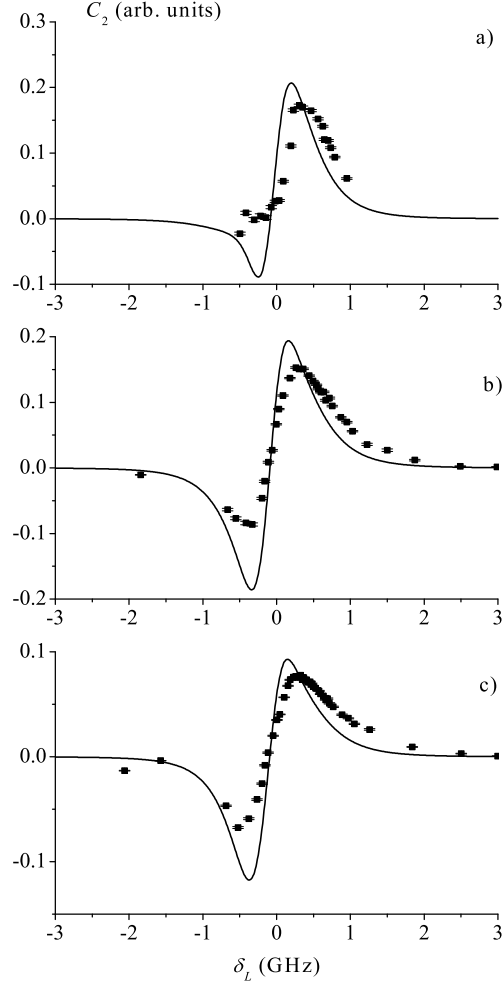}
\caption{Dispersive coefficient, $C_2$, versus optical detuning,
$\delta_L$. Plots a), b) and c) are for ${\cal R} = 2.4,\,7.2,\,22$,
respectively. The solid lines indicate the theoretical predictions while
the points indicate the experimental data.}
\label{fig4}
\end{figure}

\begin{figure}
\includegraphics[width=3in]{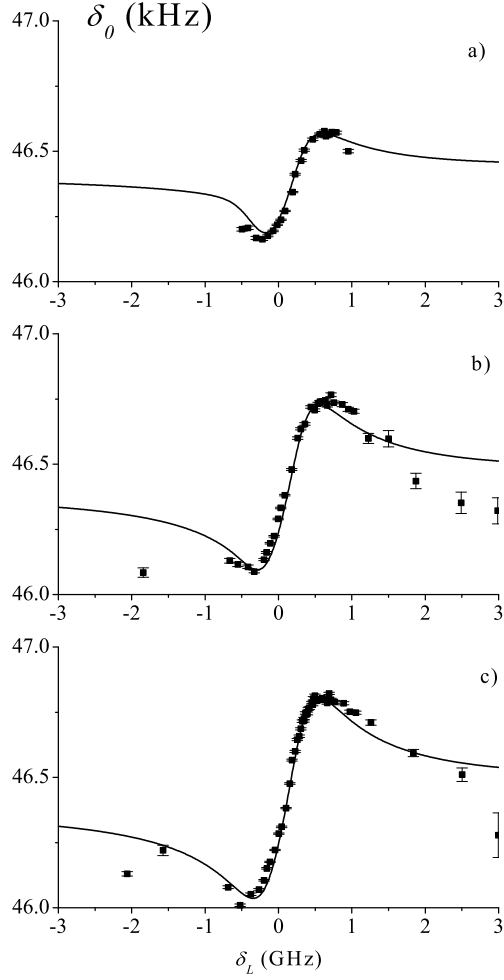}
\caption{Frequency shift, $\delta_0$, versus optical detuning,
$\delta_L$. Plots a), b) and c) are for ${\cal R} = 2.4,\,7.2,\,22$,
respectively. The solid lines indicate the theoretical predictions while
the points indicate the experimental data.}
\label{fig5}
\end{figure}

\begin{figure}
\includegraphics[width=3in]{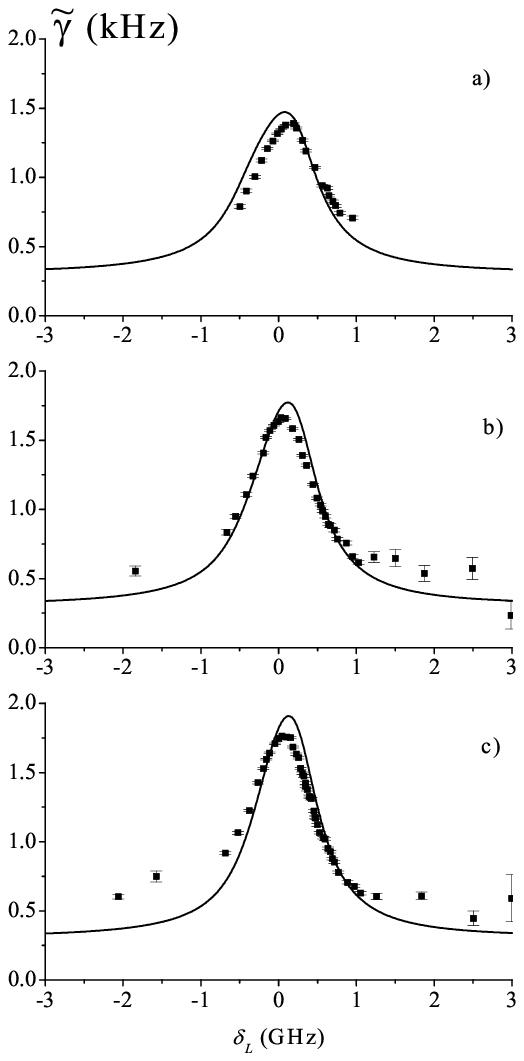}
\caption{Dark resonance width, $\widetilde{\gamma}$, versus optical
detuning, $\delta_L$. Plots a), b) and c) are for ${\cal R} =
2.4,\,7.2,\,22$, respectively. The solid lines indicate the theoretical
predictions while the points indicate the experimental data.}
\label{fig6}
\end{figure}

Since these
experimental data for Cs in Ne are fitted by \eqref{LLc} quite well,
we can compare analytically calculated coefficients of the
generalized Lorentzian to those extracted from experimental data.
The dependence of the coefficients on the total light intensity
${\cal I} \propto |E_1|^2+|E_2|^2$ is almost trivial, at least when
the power broadening $(R_1^{(m)}+R_2^{(m)})/2$ exceeds the dephasing
rate~$\Gamma$ in zero field: all the parameters $C_1$, $C_2$,
$\delta_0$, and $\widetilde{\gamma}$ scale as ${\cal I}$.  Thus, the
most representative test is provided by the dependence of the
coefficients on the one-photon detuning $\delta_L$, and  on the
intensity ratio ${\cal R} = |E_1|^2/|E_2|^2$ between the two field
components. Such comparisons with experimental fit parameters from
\cite{knappe02} are presented in Figs.~\ref{fig3} to \ref{fig6},
where $C_1$, $C_2$, $\delta_0$, and $\widetilde{\gamma}$ are plotted
as functions of $\delta_L$ for three different relative intensities,
${\cal R}$. The other parameters used in the calculations correspond
to the experimental conditions: excitation by $\sigma^{+}$ polarized
radiation, total intensity ${\cal I} = 0.4\,$mW/cm$^2$, optical
linewidth $\gamma =2\pi\, 750\,$MHz, and ground-state relaxation rate
$\Gamma = 2\pi\, 150\,$Hz. We use no free parameters, just a single
trivial scaling factor for $C_1$ and $C_2$, and a constant offset for
$\delta_0$ that accounts for the collisional shift of the dark
resonance position.

We see a good qualitative agreement, especially for the resonance
position $\delta_0$ and for the width $\widetilde{\gamma}$. There are
some noticeable discrepancies for the amplitudes $C_1$ and $C_2$. In
particular, we can see that the theoretical curve for $C_1$ can cross
the zero level at large $\delta_L$, which can be attributed to the
well-known Raman absorption, but which is not observed in the experimental data.

\section{$D_2$ line excitation and connection to previously existing
theories}

In the specific case of the $D_2$ line of Cs at high buffer-gas
pressure, the two-photon amplitudes $U$ and $V$ are much smaller than the
optical pumping rates $R^{(0)}_i$ and the optical shifts $S^{(0)}_i$,
respectively, because the most probable optical transitions $F_1 \to
F_e=I+J_e$ and $F_2 \to F_e=I-J_e$ contribute to the one-photon
transitions but not to the two-photon Raman coupling.
Note that the ratio between $V$ and $R^{(0)}_i$ can be arbitrary,
depending on the one-photon detuning $\delta_L$. As a result, the
part of the absorption signal that varies with $\delta_R$ is small
compared to the constant one, and we arrive, to lowest orders, at the
following approximate expressions. The parameter
\[
x \approx -\frac{D}{2\,C} =
\frac{2\,(\widetilde{\Gamma}_1-\widetilde{\Gamma}_2)\,U\,V}
{\widetilde{\Gamma}_1 \widetilde{\Gamma}_2}
\]
is negligible with respect to the other contributions in $\delta_0$,
$\widetilde{\gamma}$  and $C_1$. The resonance position offset and the
width are approximated as
\begin{eqnarray} \label{LL_D2}
\delta_0 &\approx& S^{(0)}_1  - S^{(0)}_2
\nonumber \\
(\widetilde{\gamma}/2)^2 &\approx& \frac{E}{C} \approx
\widetilde{\Gamma}_{12}^2 +
\frac{(\widetilde{\Gamma}_1+\widetilde{\Gamma}_2)^2}
{\widetilde{\Gamma}_1 \widetilde{\Gamma}_2}\,V^2
\;.
\end{eqnarray}
The amplitudes $C_1$ and $C_2$ are given by (\ref{c1}) and (\ref{c2})
with $x=0$ and $\widetilde{\gamma}$ from (\ref{LL_D2}).

These results can be compared with those for a  three-level $\Lambda$
system in the low-saturation limit. Our formulas \eqref{LLc}-\eqref{c2}
will describe this last case, as well, if we set $Z=0$, i.e.,
\begin{eqnarray} \label{LL_lambda}
&& \delta_0 = S^{(0)}_1  - S^{(0)}_2 \;;\;\;\;
(\widetilde{\gamma}/2)^2 = \frac{B}{A} =
\widetilde{\Gamma}_{12}^2 + 4\,V^2
\nonumber \\
&& C_1 (\widetilde{\gamma}/2)^2 = \frac{B C - A E }{A^2}
\;;\;\;\;
C_2\, \widetilde{\gamma}/2 = \frac{D}{A} \;.
\end{eqnarray}
Thus, the results are qualitatively similar (the main differences are
the overestimated amplitudes $C_1$ and $C_2$), but now all parameters
are unambiguously defined for the actual atomic structure.

When $C_2 = 0$ the lineshape is symmetrical, and occurs if $V=0$ or
$\widetilde{\Gamma}_1 = \widetilde{\Gamma}_2$. The first condition
generalizes to $\delta_L = 0$, and the second corresponds to the
condition of equal Rabi frequencies in a simple $\Lambda$ system.

When $V = 0$, the amplitude of the symmetrical signal is proportional
to the square of the two-photon coupling:
\begin{equation} \label{c1_pert_sym}
C_1 \approx \frac{2\,(\widetilde{\Gamma}_1+\widetilde{\Gamma}_2)^2}
{\widetilde{\Gamma}_{12}\,(\widetilde{\Gamma}_1+\widetilde{\Gamma}_2+Z\,
\widetilde{\Gamma}_1\widetilde{\Gamma}_2)^2}\,
U^2 \;,
\end{equation}
which is a key point of the perturbative studies \cite{wynands98} but
now, in addition, all effects of the optical pumping are accounted for in
the prefactor in \eqref{c1_pert_sym}.

\section{Dark resonance position. Three possible definitions}

The center position of the dark resonance in essence determines the
output frequency of the frequency reference or the magnetic field
indicated by the magnetometer. Especially for asymmetrical resonances,
it is somewhat unclear exactly how that center position is defined. The
quantity $\delta_0$ above is one possible definition of the resonance
position, corresponding to the combined minimum of the absorptive part,
and zero of the dispersive part, of the resonance described by
\eqref{00abs}.

\begin{figure}
\includegraphics[width=3in]{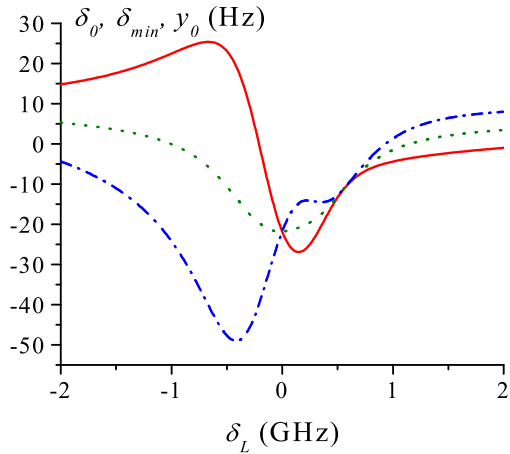}
\caption{Three possible definitions of the dark resonance position. The
centroid $\delta_0$ corresponds to the solid line, $\delta_{\mathrm
min}$ -- dotted, and $y_0$ -- dash-dotted line. All curves are
calculated for the Cs $D_2$ line. The parameters are ${\cal I} = 45
\,\mu$W/cm$^2$, ${\cal R} = 0.5$, and $\gamma = 2\pi\,850\,$MHz.}
\label{fig7}
\end{figure}

Using \eqref{LLc}-\eqref{c2}, one can easily find another possible
definition of the resonance center: the Raman detuning corresponding to
minimum absorption.
\begin{equation}  \label{dmin}
\delta_\mathrm{min} = S_{1}^{(0)}- S_{2}^{(0)} +
\frac{\widetilde{\Gamma}_{12}\,(\widetilde{\Gamma}_{2}
-\widetilde{\Gamma}_{1})}
{\widetilde{\Gamma}_{1} + \widetilde{\Gamma}_{2}} \,\frac{V}{U}
\;.
\end{equation}
A third possible definition is the point $y_0$, where the dispersion
$n'_{\mathrm{DR}}$ associated with the absorption~\eqref{LLc} (by the
Kramers-Kronig relations) is equal to zero. This is found to be
\begin{equation}  \label{y0}
y_{0} = \delta_{0} - \frac{\widetilde{\gamma}}{2}\,
\frac{C_2}{C_1}
% \approx \delta_{0} - \frac{A D}{B C- A E}\,\frac{E}{C}
\;.
\end{equation}
Each of these three quantities, $\delta_\mathrm{min}$, $y_0$ and
$\delta_0$, could be considered the resonance center, depending on how
the resonance is measured experimentally. In the general asymmetrical
case, when $V\neq 0$ (non-zero effective one-photon detuning) and
$\widetilde{\Gamma}_{1} \neq \widetilde{\Gamma}_{2}$ (unbalanced optical
pumping rates), all three values are different. Even their behavior
versus $\delta_L$ are qualitatively different (Fig.~\ref{fig7}): near
the one-photon resonance ($V=0$) the centroid $\delta_0$ of the
Lorentzians has a dispersion-like shape, while $\delta_{\rm min}$ is
rather of an absorptive nature, and $y_0$ has a more complicated shape
of mixed type. In addition, $\delta_0$ and $\delta_{\rm min}$ are always
finite, whereas $y_0$ goes to infinity at the zeros of $C_1$. These
different dependences on optical detuning could, for example, alter the
sensitivity of the frequency reference or magnetometer to the optical
lock point. As a result, careful consideration must be given to the
resonance detection method when designing frequency references or
magnetometers based on dark resonances.

\section{Conclusion}

Using very simple assumptions about the relaxation processes, analytical
results can be obtained for the nonlinear absorption of bichromatic
radiation near a two-photon resonance. The theory fully takes into
account both the HF and the Zeeman level structures of alkali atoms as
well as all light-induced effects. Our results constitute a good basis
for understanding experimental works and further possible refinements of
theory are possible. In particular, the case of large Doppler width $k\overline{v} >
\gamma$ can be immediately studied by the substitution $\delta_L \to
\delta_L -kv$ followed by averaging over the Maxwell distribution.

In addition, the theory allows for a simple parameterization of
experimentally measured dark resonances in terms of absorptive and
dispersive components. The theory can therefore predict, for example,
the detuning for which the dispersive part of the resonance is minimized
and, for a given detuning, the asymmetry in the resonance lineshape that
might be expected. The analysis of the different definitions of the
resonance center position is also of interest for practical applications
based on dark resonances such as atomic frequency standards and
magnetometers. It appears likely that the additional understanding
gained by the thorough theoretical analysis presented here will lead to
further refinement and development of current and future applications
based on dark resonances.

\begin{acknowledgments}
We thank S. Knappe, C. Affolderbach, I. Novikova, A. Matsko, and H.
Robinson for helpful discussions. AVT and VIYu were financially
supported by RFBR (grants \# 01-02-17036 and \# 03-02-16513). This
work is a contribution of NIST, an agency of the US Government, and
is not subject to copyright.
\end{acknowledgments}

\appendix

\section{Derivation of equation (\ref{maineq})}

In this appendix we consider in detail the derivation of the basic equation set (\ref{maineq}). As
is well-known the atomic density matrix obeys to the generalized optical Bloch equation. According
to this equation, the evolution of the density matrix can be split into the two parts. The
reversible one ($d/dt\,\widehat{\sigma}= -i/\hbar\,[\widehat{H},\widehat{\sigma}]$) is governed by
the total Hamiltonian of an atom in a resonant external field $\widehat{H} =
\widehat{H}_{0}+\widehat{H}_{D-E}$. The irreversible part originated from the interaction with
environments (e.g. buffer gas or vacuum modes of electromagnetic field) are modeled by relaxation
(super)operators of various kinds. The concrete form of the relaxation terms will be specified in
the course of the derivation.

The first stage is the elimination of the optical coherences $\widehat{\sigma}_{eg} =
\widehat{\Pi}_e \widehat{\sigma} \widehat{\Pi}_g$, where the operator $\widehat{\Pi}_e =
\sum_{m_e} |F_e,m_e \rangle\langle F_e,m_e|$ projects on the given HF component of the excited
state. In the low-saturation limit the optical coherence matrix obeys the following equation in
the rotating frame:
\begin{equation} \label{eg_eq}
\left[ \frac{d}{dt}+\gamma/2
-i(\delta_L-\omega_e)\right]\widehat{\sigma}_{eg} =
\frac{i}{\hbar}\, \left\{ \sum_{i=1,2}\widehat{\Pi}_e(\widehat{\bf
d }\cdot {\bf E}_i)\widehat{\Pi}_i+\sum_{i\neq
j}\widehat{\Pi}_e(\widehat{\bf d }\cdot {\bf E}_i)\widehat{\Pi}_j
e^{-i(\omega_i-\omega_j)t} \right\}\widehat{\sigma}_{gg} \;.
\end{equation}
On the left-hand side, the Raman detuning
$\delta_R$ is small compared to the homogeneous width $\gamma$
($|\delta_R| \ll \gamma$); $\widehat{\Pi}_i = \sum_{m} |F_i,m
\rangle\langle F_i,m|$, so that
$\widehat{\Pi}_g=\widehat{\Pi}_1+\widehat{\Pi}_2$. As is
explained in the main text, the
oscillations of the ground-state density submatrix
$\widehat{\sigma}_{gg}$ can also be safely neglected in the rotating frame. Then, in the
stationary regime ($\gamma t \gg 1$) the solution of the equation
(\ref{eg_eq}) is
\begin{equation} \label{eg_sol}
\widehat{\sigma}_{eg} = \frac{i}{\hbar}\, \left\{
\sum_{i=1,2}\frac{\widehat{\Pi}_e(\widehat{\bf d }\cdot {\bf
E}_i)\widehat{\Pi}_i}{\gamma/2 -i(\delta_L-\omega_e)}+\sum_{i\neq
j}\frac{\widehat{\Pi}_e(\widehat{\bf d }\cdot {\bf
E}_i)\widehat{\Pi}_j e^{-i(\omega_i-\omega_j)t}}{\gamma/2
-i(\delta_L-\omega_e)-i(\omega_i-\omega_j)}
\right\}\widehat{\sigma}_{gg} \;.
\end{equation}
Under the conditions considered here, the equation for the
ground-state density submatrix can be written
\begin{equation} \label{gg_eq}
\frac{d}{dt} \widehat{\sigma}_{gg} = -\Gamma \,
(\widehat{\sigma}_{gg}-\widehat{\sigma}^{(0)}_{gg})-\frac{i}{\hbar}[\widehat{H}_0,\widehat{\sigma}_{gg}]
-\frac{i}{\hbar}
\left(\widehat{\Pi}_g\,\overline{\widehat{H}_{D-E}\,\widehat{\sigma}}\,\widehat{\Pi}_g
 -h.c. \right)+\widehat{\cal A}\{\widehat{\sigma}_{ee}\} \;,
\end{equation}
where the line over operators indicates time averaging, i.e. all
the oscillating terms should be removed from the product
$\widehat{H}_{D-E}\,\widehat{\sigma}$. Using (\ref{eg_sol}), one
finds that
\[
-\frac{i}{\hbar}
\widehat{\Pi}_g\,\overline{\widehat{H}_{D-E}\,\widehat{\sigma}}\,\widehat{\Pi}_g
= - i\, \widehat{R}\,\widehat{\sigma}_{gg}\;
\]
where $\widehat{R}$ is the excitation matrix given by (\ref{Rmatr}). The
first term on the right-hand side of (\ref{gg_eq})
describes the relaxation in the ground state (due to both
diffusion and collisions) toward the equilibrium distribution outside
the laser beam, $\widehat{\sigma}^{(0)}_{gg} =
\widehat{\Pi}_g/n_g$. All the linear (with respect to
$\widehat{\sigma}_{gg}$) terms, containing $\Gamma$,
$\widehat{H}_0$, and $\widehat{R}$, can be combined in the
effective non-Hermitian Hamiltonian \eqref{Heff}. The last term on
the right-hand side of (\ref{gg_eq}) corresponds to the
spontaneous radiative transfer of atoms from the
excited-states, given by the density submatrix
$\widehat{\sigma}_{ee}=\widehat{\cal
P}_e\,\widehat{\sigma}\,\widehat{\cal P}_e$ (where $\widehat{\cal
P}_e=\sum_{F_e}\widehat{\Pi}_e$), to the ground-state levels. Its
structure will be specified below.

In the low-saturation limit, the matrix $\widehat{\sigma}_{ee}$
obeys the equation
\begin{equation} \label{ee_eq}
\frac{d}{dt} \widehat{\sigma}_{ee} = -\frac{1}{\tau_e} \,
\widehat{\sigma}_{ee}-\frac{i}{\hbar}[\widehat{H}_e,\widehat{\sigma}_{ee}]
-\widehat{\cal G}\{\widehat{\sigma}_{ee}\} -\frac{i}{\hbar}
\left(\widehat{\cal
P}_e\,\overline{\widehat{H}_{D-E}\,\widehat{\sigma}}\,\widehat{\cal
P }_e
 -h.c. \right) \;,
\end{equation}
where the first three terms on the right-hand side describe the
radiative decay, the HF splitting ($\widehat{H}_e =
\hbar\,\sum_{F_e}\omega_e\,\Pi_e$), and the collisional
depolarization of the excited state, respectively; the last term
corresponds to the excitation due to light-induced transition from
the ground-state levels. This last term can be considered as a
source, because it is proportional to $\widehat{\sigma}_{gg}$:
\begin{eqnarray*}
-\frac{i}{\hbar} \widehat{\cal
P}_e\,\overline{\widehat{H}_{D-E}\,\widehat{\sigma}}\,\widehat{\cal
P }_e &=& \frac{1}{\hbar^2}\sum_{F_e,F_e'}\left( \sum_{i,j}
\frac{\widehat{\Pi}_{e'}(\widehat{\bf d}\cdot {\bf E}_i)
\widehat{\Pi}_i\widehat{\sigma}_{gg} \widehat{\Pi}_j
{(\widehat{\bf d}\cdot {\bf E}_j)}^{\dagger}
\widehat{\Pi}_{e}}{\gamma/2+i(\delta_L-\omega_e)}+\right.\\
&+&\left. \sum_{i\neq j} \frac{\widehat{\Pi}_{e'}(\widehat{\bf
d}\cdot {\bf E}_i) \widehat{\Pi}_j\widehat{\sigma}_{gg}
\widehat{\Pi}_j {(\widehat{\bf d}\cdot {\bf E}_i)}^{\dagger}
\widehat{\Pi}_{e}}{\gamma/2+i(\delta_L-\omega_e+\omega_i-\omega_j)}
 \right) \;.
\end{eqnarray*}
The structure of the collisional term $\widehat{\cal G}\{\widehat{\sigma}_{ee}\}$ can be found in
\cite{happer72}. Here we simply recall that during the course of a collision only the electronic
component of the atomic polarization is depolarized. The nuclear component is involved in the
process of depolarization due to the HF coupling. For all alkali atoms, the excited-state HF
splitting $\Delta_e$ is much greater than radiative decay rate $1/\tau_e$. In addition we assume
that the collisional relaxation rates $\gamma_{\kappa}$ for the excited-state electronic multipole
moments of rank $\kappa=1,\ldots,2J_e+1$ also obey the conditions $\gamma_{\kappa} \tau_e \gg 1$
(for $\kappa=0$ we assume $\gamma_0=0$, i.e. the collision-induced transitions between the fine
structure components are not considered here). In this limit, $\Delta_e \tau_e \gg 1$ and
$\gamma_{\kappa} \tau_e \gg 1$, the steady-state solution of \eqref{ee_eq} has particularly simple
form:
\begin{equation} \label{ee_sol}
\widehat{\sigma}_{ee} = \pi_e\,\frac{\widehat{\cal
P}_e}{n_e}\;;\;\;\;\; \pi_e =\tau_e\left(i\,{\rm
Tr}\{\widehat{R}\,\widehat{\sigma}_{gg}\}+c.c.\right) \;,
\end{equation}
which corresponds to total collisional depolarization of the
excited state.

Here we shall illustrate this fact in one specific case, when the
excited-state HF splitting is much larger than the depolarization
rates $\gamma_{\kappa}$ and when all the depolarization rates
(except for $\gamma_0$) are the same (so-called pure
electronic randomization model \cite{happer72}). If $\Delta_e \gg
\gamma_{\kappa},\,1/\tau_e$, one can neglect HF coherence in the
excited state. For pure electronic randomization both eigenvalues
and eigenvectors of the Liouvillian ${\cal G}$ are well-known
\cite{happer72}, which allows us to write the steady-state
solution of \eqref{ee_eq} for arbitrary $\gamma_{\kappa} \tau_e$:
\begin{eqnarray} \label{ee_sol_per}
\widehat{\sigma}_{ee} &=& \frac{\tau_e}{1+\gamma_{\kappa}\tau_e}
\widehat{S}_{e} +
\frac{\gamma_{\kappa}\tau_e}{1+\gamma_{\kappa}\tau_e}
\sum_{L,M,F_e,F_e'}\frac{\tau_e}{1+ \widetilde{\gamma}_L
\tau_e}(-1)^{F_e-F_e'} \frac{(2 F_e+1)(2 F_e'+1)}{(2 J_e+1)}
\times \nonumber
\\ &\times& \left\{
\begin{array}{rcl}
F_e & F_e & L\\ I & I & J_e
\end{array}
\right\} \left\{
\begin{array}{rcl}
F_e' & F_e' & L\\ I & I & J_e
\end{array}
\right\} \widehat{T}_{LM}(F_e\,F_e) {\rm
Tr}\{\widehat{T}_{LM}^{\dagger}(F_e'\,F_e')\widehat{S}_{e}\} \;.
\end{eqnarray}
Here the source has the form
\begin{eqnarray*}
\widehat{S}_e &=& \frac{\gamma}{\hbar^2}\sum_{F_e}\left(
\sum_{i,j} \frac{\widehat{\Pi}_{e}(\widehat{\bf d}\cdot {\bf E}_i)
\widehat{\Pi}_i\widehat{\sigma}_{gg} \widehat{\Pi}_j
{(\widehat{\bf d}\cdot {\bf E}_j)}^{\dagger}
\widehat{\Pi}_{e}}{(\gamma/2)^2+(\delta_L-\omega_e)^2}+\right.\\
&+&\left. \sum_{i\neq j} \frac{\widehat{\Pi}_{e}(\widehat{\bf
d}\cdot {\bf E}_i) \widehat{\Pi}_j\widehat{\sigma}_{gg}
\widehat{\Pi}_j {(\widehat{\bf d}\cdot {\bf E}_i)}^{\dagger}
\widehat{\Pi}_{e}}{(\gamma/2)^2+(\delta_L-\omega_e+\omega_i-\omega_j)^2}
 \right) \; ;
\end{eqnarray*}
the relaxation rates
\begin{equation} \label{gammaL}
\widetilde{\gamma}_L = \gamma_{\kappa} \left[
1-\sum_{F_e}\frac{(2F_e+1)^2}{(2J_e+1)} \left\{
\begin{array}{rcl}
F_e & F_e & L\\ I & I & J_e
\end{array}
\right\}^2 \right] \;;\;\;\;\;\; L=0,\ldots ,2I+1
\end{equation}
correspond to the Zeeman projections of the nuclear multipole
moments of rank~$L$ \cite{happer72}; and the Wigner tensorial
operators are defined as
\[
\widehat{T}_{LM}(F_a\,F_b) = \sum_{m_a,m_b} |F_a,m_a\rangle \,
\sqrt{2L+1}\,(-1)^{F_a-m_a} \left(
\begin{array}{rcl}
F_a & L & F_b\\ -m_a & M & m_b
\end{array}
\right)\, \langle F_b,m_b| \;.
\]
As is seen from \eqref{gammaL} the rates $\widetilde{\gamma}_L$
are of the order of $\gamma_{\kappa}$ apart from
$\widetilde{\gamma}_0 =0 $. Then in the limit $\gamma_{\kappa}
\tau_e \gg 1$ the leading term of \eqref{ee_sol_per} corresponds
to the summand with $L=0$, which leads directly to the solution
\eqref{ee_sol}.

When the excited-state HF coherence is negligible, the radiative
repopulation term in \eqref{gg_eq} can be written as
\begin{equation} \label{A}
\widehat{\cal A}\{\widehat{\sigma}_{ee}\} =
\frac{1}{\tau_e}\sum_{F_e,i,q}\frac{r(F_e,F_i)^2}{3}\,
\widehat{T}_{1q}^{\dagger}(F_e\,F_i)\, \widehat{\sigma}_{ee} \,
\widehat{T}_{1q}(F_e\,F_i) \; .
\end{equation}
One can easily prove the fundamental property:
\begin{equation} \label{isotropy}
\widehat{\cal A}\{\widehat{\cal P}_{e}\} =
\frac{1}{\tau_e}\frac{n_e}{n_g}\, \widehat{\Pi}_{g} \;,
\end{equation}
which expresses the isotropy of the radiative relaxation.

Thus, we see that in the case of total collisional depolarization
of the excited state, when the excited-state density matrix is
proportional to $\widehat{\cal P}_{e}$ [as shown in \eqref{ee_sol}],
\eqref{gg_eq} is reduced to \eqref{maineq}. In addition,
the expression for the optical coherence matrix \eqref{eg_sol}
allows one to calculate various spectroscopic signals (as well as
the total absorption), for example, the total dispersion.

%\bibliography{lorlor2}

\end{document}